\documentclass[prd,nofootinbib,aps]{revtex4}

\usepackage{graphicx}

\begin{document}
\title{Undecidability and the problem of outcomes in quantum measurements}

\author{Rodolfo Gambini$^{1}$,
Luis Pedro Garc\'{\i}a Pintos${^1}$,
and Jorge Pullin$^{2}$}
\affiliation {1. Instituto de F\'{\i}sica,
Facultad de Ciencias, Igu\'a 4225, esq. Mataojo, Montevideo, Uruguay. \\
2. Department of Physics and Astronomy, Louisiana State
University, Baton Rouge, LA 70803-4001}

\date{May 23rd 2009}

\begin{abstract}
We argue that it is fundamentally impossible to recover information
about quantum superpositions when a quantum system has interacted with a
sufficiently large number of degrees of freedom of the
environment. This is due to the fact that gravity imposes fundamental
limitations on how accurate measurements can be. This leads to the
notion of undecidability: there is no way to tell, due to fundamental
limitations, if a quantum system evolved unitarily or suffered
wavefunction collapse.  This in turn provides a solution to the
problem of outcomes in quantum measurement by providing a sharp
criterion for defining when an event has taken place.  We analyze in
detail in examples two situations in which in principle one could
recover information about quantum coherence: a) ``revivals'' of
coherence in the interaction of a system with the measurement
apparatus and the environment and b)
the measurement of global observables of the system plus apparatus
plus environment. We show in the examples that the fundamental
limitations due to gravity and quantum mechanics in measurement
prevent both revivals from occurring and the measurement of global
observables.  It can therefore be argued that the emerging picture
provides a complete resolution to the measurement problem in quantum
mechanics.
\end{abstract}
\maketitle

\section{Introduction}

The Copenhagen interpretation of quantum mechanics is formulated with
the aid of a classical macroscopic realm in order to explain the
measurement process. Given the growing number \cite{experiments} of
experiments showing the existence of superpositions of macroscopically
distinct quantum states, it is becoming increasingly desirable to
understand quantum mechanics entirely as a standalone quantum paradigm
without having to refer to an external classical world (see for
instance \cite{omnes,schlossauer} for previous attempts).  Notice
that the problem transcends the actual physical measurement of
quantities in that it is present every time one wishes to assign a
probability to an event. We use the word event to denote that a
physical observable of coupled system plus apparatus plus environment
takes a definite value. (For more details on the notion of event in
light of the ideas of this paper see appendix 2.)

Most physicists claim that environmentally induced decoherence may
play an important role in the solution of the measurement problem of
quantum mechanics (see \cite{schlossauer} for a recent review).
Decoherence allows us to understand how the interaction with the
environment induces a local suppression of interference between a set
of preferred states, associated with the ``pointer basis''. It leads
to a fast suppression of the interference terms of the reduced matrix
describing the system $S$ coupled to the measurement apparatus $A$,
and it selects a preferred set of states that are robust in spite of
their interaction with the environment. All this is achieved with the
standard unitary time evolution for the composite system plus
apparatus plus environment $S+A+E$ and therefore the global phase
coherence is not destroyed but simply transferred from the system $S$ to
the environment.

There are two main criticisms to the solution of the problem of
measurement by invoking the environment (decoherence). The first
criticism is that that the ``system plus apparatus plus environment''
evolves unitarily and therefore all information about the original
superposition is still present at the end and could in principle be
retrieved (see for instance
\cite{despagnat}). The second type of criticism is related to the
fact that the system plus the measurement apparatus
is left in a superposition of states through the
interaction with the environment  and therefore it would not generate a
definite event (or measurement) but a superposition of them.
Bell \cite{bell} calls this  ``and/or problem'', since the states corresponding
to the diagonal of the density matrix coexist with each other and
therefore one has ``A and B'' as possibilities. In the classical world,
however, one has ``A or B'' as the resulting outcomes.

In previous work \cite{foundations} we have argued that fundamental
limitations to the measurement of time and distances due to
gravitational effects imply that there are no mechanisms to retrieve
the information about a quantum superposition from the system plus
apparatus plus environment.  Essentially, gravity puts fundamental
limitations on the accuracy of clocks and this adds a new source of
loss of quantum coherence.  This in turn implies that experiments that
``wait a long time'' for quantum coherence to be restored
(``revivals'') \cite{despagnat} actually worsen the situation due to
the increasing inaccuracy of clocks as one measures longer times (see
\cite{foundations} for details). A second proposal to recover quantum
coherence, by measuring global observables of the ``system plus
measuring apparatus plus environment'' is also limited by fundamental
arguments based on the laws of quantum mechanics and gravity, as we
shall see.

If one accepts that quantum coherence cannot be retrieved, this leads
to another key point we have recently emphasized, undecidability
\cite{undecidability}:  one cannot decide if the
quantum state has suffered reduction or if it evolved unitarily. In
fact, it could even be conceivable that sometimes there might be
reduction, sometimes not, and we do not have reasons to expect one or
the other in a given instance.  Let us get back to the second
objection to the decoherence solution to the measurement problem: that
at the end of the interaction with the environment, the measuring
apparatus is generically left in a superposition of (eigen)-states
corresponding to different ``positions of the needle of the
gauge''. That would not correspond to what one usually calls a
``measurement'' (production of an event) in which the apparatus is in
a given (eigen-)state, corresponding to the ``needle of the gauge''
taking a definite position. This is the and/or problem of Bell that we
mentioned above.  We would like to argue that this problem can be
solved by defining the appearance of events without necessarily
implying a change in the quantum state. We will claim that an event
occurs when the distinction between the ``system plus apparatus plus
environment'' being in a superposition or in a given state becomes
{\em undecidable}. At this point a definite choice of outcome, be it A
or B, is compatible with the laws of physics. All possible experiments
would give the same results if the complete system plus apparatus
plus environment suffered a reduction of its state or evolved
unitarily.

The organization of this paper is as follows. In section II we briefly
review the fundamental limitations that gravity imposes on the
measurement process and how it leads to a fundamental loss of
coherence in quantum states. In section III we review how the
limitations on measurement lead to a notion of undecidability in which
one cannot tell if a quantum collapse has occurred or not and how this
provides a sharp criterion for when a quantum event took place. In
section IV we analyze the proposal in the context of models of
environmentally-induced decoherence related to the Zurek model of
spins. In particular we show how undecidability occurs and how the
problem of ``revivals'' and the ``problem of outcomes'' are solved.
In section V we analyze if extreme physics in quantum field theory
could violate the results of section IV. We end with a discussion.

\section{Fundamental loss of coherence due to the use of real clocks and measuring rods}

Quantum mechanics as ordinarily formulated has a unitary evolution. In
the usual formulation one assumes that time $t$ is given by a
classical parameter. This parameter can be measured with arbitrary
precision. (A similar assumption is made about space if measurements
if one is dealing with quantum field theory). But one knows that no
observable quantity in nature can be measured with absolute
precision. At the very least one will have quantum fluctuations in the
observable. But theoretically one could choose observables with
arbitrarily small quantum fluctuations and use those as the
``clock''. This hits a fundamental limit if one includes gravity in
measurements. To make accurate measurements one need to expend
energy. When one considers gravity one is limited in the amount of
energy one can invest in a measurement.  Too large an energy density
creates a black hole and spoils the measurement.  This argument has
been put forward by many authors. It is heuristic and without a full
theory of quantum gravity cannot be rigorously worked out. The
heuristic limits all imply that the error in the measurement of a time
$T$ goes as $\delta T\sim T_{\rm Planck}^{1-a} T^{a}$ with $T_{\rm
Planck} \sim 10^{-44}s$ is Planck's time and $a$ is a positive power.
Some variations of the argument yield $a=1/3$, some yield $a=1/2$
\cite{amelino}. The particular value of $a$ is not very relevant, as
we shall discuss later. We will choose $a=1/3$ for the rest of the
paper since the value is suggested by several arguments
(\cite{ng,lloyd,karolyhazy}). As long as the error in measuring the
time goes as a positive power of $T$ then one loses quantum
coherence. The reason for this is simple: although quantum evolution
is unitary in terms of the fiduciary parameter $t$ that appears in the
Schr\"odinger equation, our clocks are not good enough to keep track
of unitarity in evolution. After some time has evolved, even if we
started with a pure state, our inaccurate clocks will force us to
consider a superposition of states at different values of $t$ as
corresponding to the value of ``time'' $T$.  Therefore pure states
evolve into mixed states. A similar argument can be worked out for the
measurements of distance in the context of quantum field theory but we
will not expand it here, readers can refer to \cite{spatial} for more
details.

The loss of coherence due to imperfect clocks makes the off diagonal
elements of the density matrix of a quantum system in the energy
eigenbasis to decrease exponentially. The exponent for the $mn$-th
matrix element is given by $\omega_{mn}^2 T_{\rm Planck}^{4/3}
T^{2/3}$, where $\omega_{mn}=E_{mn}/\hbar$ is the difference of energy
between levels $m$ and $n$ divided by $\hbar$ (the Bohr frequency
between $n$ and $m$). One could see this effect in the lab
in reasonable times (hours) only if one were handling ``macroscopic''
quantum states corresponding to about $10^{13}$ atoms in coherence.
The direct observation of this effect is therefore beyond our current
experimental capabilities. However, as we shall see, it has important
implications for the measurement problem in quantum mechanics.

\section{Undecidability as the criterion for the production of events}

The ``problem of outcomes'' associated with the measurement problem in
quantum mechanics has to do with the fact that even if one assumes the
quantum system interacts with an environment and decoherence takes
place, and the density matrix of the system (plus the measurement
apparatus if there is one present) is diagonal, the density matrix
represents an improper mixture and therefore the quantum system (plus
the measurement apparatus)  is generically left in a superposition of
states. Yet, when a measurement occurs one usually assumes the system
 and the measurement apparatus are  left in an individual, definite
state, not in a superposition. Therefore it appears that decoherence
alone cannot account for the measurement process in quantum mechanics.

There have been several proposals in the literature to address this
problem.  One type of proposals consists in modifying the dynamics of
the quantum theory to actually drive the system to an individual
definite state.  Examples of this point of view would be the modified
dynamics of Ghirardi, Rimini and Weber \cite{grw} or Penrose
\cite{Penrose}.  Another point of view is to consider that the
evolution is given by the traditional, unitary, Schr\"odinger
equation, and to associate a criterion to decide which are the
definite-valued observables, and therefore which are the events that
may occur. This type of proposals may be generically called as
``modal'' interpretations. The many worlds interpretation can also be
considered among this group. Particular proposals differ in the
criteria chosen to determine which event occurs.  A discussion of
various modal interpretations and some of their issues can be seen in
\cite{castagnino}.

We here propose a new point of view. We will give a criterion for the
production of events that is compatible both with a unitary evolution
and with a reduction postulate. We will claim that an event has
occurred when the evolution of the  composite system plus apparatus
plus environment reaches a point such that we cannot determine if a
collapse or a unitary evolution has taken place.  Usually this will
happen when a quantum system interacts with an environment i.e. it
does not occur in quantum systems in isolation. In such case a unitary
evolution or an abrupt change as the one given by a collapse would be
obviously distinguishable.  When any system reaches this state, we
will say that the system has become ``undecidable''. The appearance of
undecidability is therefore the criterion for the production of an
event. An event occurs when the system plus apparatus plus environment
becomes undecidable. We shall see that the loss of coherence due to
the use of realistic clocks and measuring rods we mentioned in the
previous section, together with the usual uncertainties of quantum
mechanics, is what is at the root of the undecidability, and is
therefore what enables the new point of view presented in this paper.

To illustrate how this would work in practice in what follows we would like
to consider a concrete example.

\section{Models of environmentally induced decoherence}

\subsection{Zurek's model}

As is customary in studies of environmental decoherence in quantum
mechanics these types of ideas cannot be proved in general, but have to be explored
in examples. Therefore in spite of the universality of the idea of
undecidability, there is of course no way to reach conclusions in
generality, it must be studied in specific examples of increasing
level of realism. The example we wish to consider is a more realistic
version of a model introduced by  Zurek
\cite{zurek} to probe ideas of environmentally-induced decoherence.
Let us start by recalling that model.  Although it does not include
all the effects of a realistic situation, it exhibits how the
information is transferred from the measuring apparatus to the
environment. It consists of taking a spin one-half system that encodes
the information about the microscopic system plus the measuring
device. A basis in its two dimensional Hilbert space will be denoted
by $\{\vert+\rangle,\vert-\rangle\}$. The environment is modeled by a
bath of many similar two-state systems called atoms. There are $N$ of
them, each denoted by an index $k$ and with associated two dimensional
Hilbert space $\{\vert+\rangle_k,\vert-\rangle_k\}$. The dynamics is
very simple, when there is no coupling with the environment the two
spin states have the same energy, which is taken to be $0$.  All the
atoms have zero energy as well in the absence of coupling. The whole
dynamics is contained in the coupling, given by the following
interaction Hamiltonian
\begin{equation}
H_{\rm int} = \hbar \sum_k \left( g_k \sigma_z \otimes \sigma_z^k
\otimes_{j\neq k} I_j\right).
\end{equation}
$\sigma_z$ is  a Pauli spin matrix acting on the state of the system.  It
has eigenvalues $+1$ for the spin eigenvector $\vert+\rangle$ and $-1$ for
$\vert-\rangle$. The operators $\sigma^k_z$ are similar, acting on the
state of the $k$-th atom. $I_j$ denotes the
identity matrix acting on atom $j$ and $g_k$ is the coupling constant
that has dimensions of frequency and characterizes the coupling energy of
one of the spins $k$ with the system. In spite of the abstract
character of the model, it can be thought of as providing a sketchy
model of a photon propagating in a polarization analyzer.

Starting from a normalized initial state
\begin{equation}
\vert\Psi(0)\rangle = \left(a\vert+\rangle + b\vert-\rangle\right) \prod_{k=1}^N \otimes \left[
\alpha_k\vert+\rangle_k +\beta_k \vert-\rangle_k \right],
\end{equation}
it is easy to solve the Schroedinger equation and one gets for the
state at the time $t$,
\begin{eqnarray}\label{3}
\vert\Psi(t)\rangle
&=& a \vert+\rangle \prod_{k=1}^N\otimes\left[
\alpha_k\exp\left(ig_k t\right)\vert+\rangle_k
+ \beta_k \exp\left(-ig_k t\right)\vert-\rangle_k\right]\\
&&+ b \vert-\rangle
\prod_{k=1}^N\otimes\left[
\alpha_k\exp\left(-ig_k t\right)\vert+\rangle_k
+ \beta_k \exp\left(ig_k t\right)\vert-\rangle_k\right].\nonumber
\end{eqnarray}

Writing the complete density operator $\rho(t) = \vert\Psi(t)\rangle\langle\Psi(t)\vert$,
one can take its trace over the environment degrees of freedom to get
the reduced density operator,
\begin{equation}
\rho_c(t) = \vert a\vert^2 \vert+\rangle\langle+\vert + \vert b\vert^2 \vert-\rangle\langle-\vert + z(t) ab^* \vert+\rangle\langle-\vert
+z^*(t) a^* b\vert-\rangle\langle+\vert,
\end{equation}
where
\begin{equation}\label{5}
z(t) = \prod_{k=1}^N \left[\cos\left(2g_k t\right)+i\left(\vert\alpha_k\vert^2
-\vert\beta_k\vert^2\right) \sin\left(2 g_k t\right)\right].
\end{equation}

The complex number $z(t)$ controls the value of the non-diagonal
elements. If this quantity vanishes the reduced density matrix
$\rho_c$ would correspond to a totally mixed state.  Ignoring the
``problem of outcomes'' for a minute, that form of the matrix is the
desired result, one would have several classical outcomes with their
assigned probabilities. However, although the expression we obtained
vanishes quickly assuming the $\alpha$'s and $\beta$'s take random
values, it behaves like a multiperiodic function, i.e. it is a
superposition of a large number of periodic functions with different
frequencies. Therefore this function will retake values arbitrarily
close to the initial value for sufficiently large times (in closed
systems). This implies that the apparent loss of information about the
non-diagonal terms reappears if one waits a long enough time. This
problem is usually called ``recurrence of coherence'' or ``revival of
coherence''.  The characteristic time for these phenomena is
proportional to the factorial of the number of involved
frequencies. This time is usually large, perhaps exceeding the age of
the universe ---at least for sufficiently large systems---, making the
problem unobservable at a fundamental rather than just a practical
level.  We will address in section \ref{Va} how the fundamental limits
on the measurements of time and distance completely eliminates the
presence of revivals for sufficiently large systems, therefore
confirming one cannot recover quantum coherence even in principle.

Another way of distinguishing if a reduction has taken place has been
proposed by d'Espagnat. His proposal consists in analyzing a global
observable for which the predictions of decoherence differ from those
of a reduction postulate.  The observable $\hat{M}$ consists in
measuring the $x$ component of the spin of the combined system plus
environment, $\hat{M}=
\hat{\sigma}_x^S
\otimes \prod_{k=1}^N \hat{\sigma}_x^k$. This observable commutes with
the model's Hamiltonian and is therefore a conserved quantity that
generically will have a non-zero expectation value. If we think in
terms of collapse, the spin has been measured and is therefore in the
state $\vert +\rangle$ or $\vert -\rangle$. The expectation value of
the observable can be shown to vanish. If we measure the observable
several times we can obtain its expectation value and check if
collapse has taken place or not. Although in practice, for this example
and for other systems, it is unrealistic to expect that one will be
able to measure observables like this one, one can ask if at least
in principle they could not be observed.

\subsection{A modified Zurek model with a more realistic interaction}

Since we have argued that Zurek's model  is based on an idealization of the
interaction with the environment, we would like to propose a model
with a more realistic interaction between the spins. This could allow
to study the behavior of d'Espagnat's observable in a more realistic
setting. The model we propose is a cavity with a uniform magnetic
field in the $z$ direction and in which there is a static spin $S$
that represents the ``needle'' of the measuring apparatus. We assume a
flux of spins (``the environment'') is pumped into the cavity and the
spins interact with $S$ during a finite time $\tau$ (the time will
depend on the speed at which the spins are pumped into the cavity and
the length of it. We can avoid having interactions among the spins of
the ``environment'' by making the stream sufficiently ``dilute'' (i.e.
essentially the spins go through the cavity one at a time).  We will
propose a more realistic interaction Hamiltonian than Zurek's.  Assume
$\gamma_1$ is the magnetic moment of the spin $S$ and $\gamma_2$ the
magnetic moment of the spins of the environment and $f_k$ the coupling
between the $k$-th spin of the environment and $S$. The interaction
Hamiltonian we choose is,
\begin{equation}
\hat{H}^{\rm int}_k = f_k \left(
\hat{S}_x \hat{S}_x^k +
\hat{S}_y \hat{S}_y^k +
\hat{S}_z \hat{S}_z^k\right).
\end{equation}
The Hamiltonian due to the presence of the magnetic field when the $k$-th particle is in
the cavity is,
\begin{equation}
\hat{H}^B_k = \gamma_1 B \hat{S}_z\otimes \hat{I} _k +\gamma_2
B\hat{I}\otimes S^{k}_z,
\end{equation}
where $\hat{I}$ is the identity matrix acting on the Hilbert space of
the needle and $I_k$ is the identity in the Hilbert space of the
$k$-th particle. The introduction of a constant magnetic field
pointing in a given direction, which we choose to be the
$z$-direction, is in order to have a definite pointer basis. Recall
that pointer states are distinguished by their ability to persist in
spite of environmental monitoring.  The system will therefore be
quasi-diagonal in the basis associated with the direction $z$. Paz and
Zurek have studied in detail the which pointer basis arises depending
on which term in the Hamiltonian is the dominant one
\cite{pazurek}. We here assume the dominant term is the one giving the
coupling of the magnetic field with the needle.  We would like to see
if we can either measure global observables or observe revivals for
this system (the needle).

The complete interaction Hamiltonian (acting on the Hilbert space of all
particles) when the particle $k$ is in the
cavity is,
\begin{eqnarray}\label{hint}
\hat{H}_k &=&\left(\hat{H}^{\rm int}_k+\hat{H}^B_k\right)\otimes \prod_{j\neq k}
\hat{I}_j \nonumber\\
&=&\left[f_k\left(\hat{S}_x\hat{S}_x^k+\hat{S}_y\hat{S}_y^k+\hat{S}_z\hat{S}_z^k\right)
+\gamma_1 B \hat{S}_z\otimes \hat{I}_k +\gamma_2 B\hat{I}\otimes \hat{S}_z^k\right]\otimes
\prod_{j\neq k} \hat{I}_j.
\end{eqnarray}
We are not explicitly writing the Hamiltonian describing the spatial evolution of the particles
of the environment, which we assume to be a free-particle Hamiltonian.

\begin{figure}
\includegraphics[height=5cm]{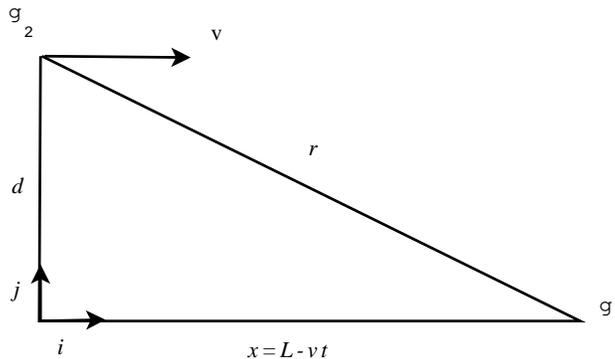}
\caption{The geometry of the interaction.}
\label{fig1}
\end{figure}

We will first analyze qualitatively the possibility of measuring the
global observable.  We make the particles traverse the cavity where
they interact with the magnetic field and the spin representing the
``needle''. One could try to measure the spins after they exit the
cavity using a Stern--Gerlach apparatus, and use the results to
compute the observable $\hat{M}$ or a similar observable involving all
the spins. Repeating the experiment for an ensemble one could compute
the expectation value $\langle \hat{M}\rangle$ and therefore determine
if collapse has taken place or not. Let us analyze the spin
coupling. The coupling constant is given by $f=\mu\gamma_1
\gamma_2/\hbar r^3$ where $\mu$ is the vacuum permeability,
$\gamma_{1,2}$ are the magnetic moments of the spins, $r$ is the
separation of the spins.  In order for the system to experiment
environmental decoherence the couplings between the spins cannot be
arbitrarily small.  For instance, if we reworked the state (\ref{3})
for this model, one gets,
\begin{eqnarray}
\vert\Psi(t)\rangle
&=& a \vert+\rangle \prod_{k=1}^N\otimes\left[
\alpha_k\exp\left(i\int dt f_k \right)\vert+\rangle_k
+ \beta_k \exp\left(-i\int dt f_k \right)\vert-\rangle_k\right]\\
&&+ b \vert-\rangle
\prod_{k=1}^N\otimes\left[
\alpha_k\exp\left(-i\int dt f_k \right)\vert+\rangle_k
+ \beta_k \exp\left(i\int dt f_k \right)\vert-\rangle_k\right].\nonumber
\end{eqnarray}
and if the interaction is too weak we would get products of quantities with
modulus close to one and the off diagonal terms of the density matrix (\ref{5}) would not cancel.
Therefore in order to have environmental decoherence the condition is $\int dt f_k > 1$.
To compute the integral we refer the reader to figure \ref{fig1},
\begin{equation}
\int f_k dt = \frac{\mu\gamma_1\gamma_2}{\hbar} \int_0^\tau \left(d^2+\left(L-vt\right)^2\right)^{-\frac{3}{2}} dt =
\frac{2 \mu \gamma_1 \gamma_2}{\hbar v d^2} \frac{1}{\sqrt{1 +\frac{d^2}{L^2}}},
\end{equation}
where $d$ is the impact parameter, $v$ is the velocity of the spins of
the environment and $2L$ is the length of the cavity.
So we can conclude that for the existence of environmental decoherence due to the interaction
with the spins one must satisfy the condition,
\begin{equation}
\frac{\mu\gamma_1\gamma_2}{\hbar d^2 v} > 1,\label{bound}
\end{equation}
and we therefore see that one either needs large values for the magnetic moments or
a small separation or velocity. However,  there is an interaction acting on the spins, its energy
of interaction is
$E \sim -\mu \gamma_1 \gamma_2/r^3$, and this leads to a
force with a component in the $y$ direction,
\begin{equation}
F_y = -\frac{2 \mu \gamma_1 \gamma_2}{r^5} y,\label{force}
\end{equation}
which causes to an acceleration in the point of closest approach
between the spins $a_y=-\mu\gamma_1\gamma_2/(md^4)$.  If the relative
speed with respect to the environment particles is $v$ then we have a
period of time of at least $d/v$ in which they will remain at a
distance of the order of $d$ from the spin $S$, being influenced by
the repulsive force. This gives us a lower bound to the speed that the
particles will acquire in the direction perpendicular to the motion of
$v_y > a_y d/v \sim
\mu \gamma_1 \gamma_2 (m d^3 v)$. Coupling this with the bound
(\ref{bound}) one gets
\begin{equation}
v_y> \hbar/(md), \label{vy}
\end{equation}
which for instance, would yield $v_y> 10^{-7}/d$ in MKS units, if the
spins of the environment are neutrons. We will see  that the
force responsible for this variation in the velocity,
together with the uncertainty in the position of the particles
in the incident package will make impossible to determine the final
position with enough precision to measure the spins of the particles
of the environment after they finish passing through the cavity. To
show this let us figure out the minimum time it will take to perform
the experiment. We see that equation (\ref{bound})  gives us a
bound on the time of flight since $\tau > d/v$. We then have $\tau >
\hbar d^3/(\mu\gamma_1 \gamma_2)$. If we take as the spin $S$ a
proton and the environment spins to be neutrons $m\sim 10^{-27}kg$,
$\gamma_{1,2} \sim 10^{-26} {\rm Joule/Tesla}$, we have that the total
time of the experiment $T>N \tau \sim N d^3 10^{25}$.  The shortest
time for the experiment is characterized by the separation of the
particles of the environment. We will take such separation to be much
larger than the smallest possible $d$, let us say $d\sim
10^{-13}m$. If, for instance, the number of particles in the
environment is $N\sim 10^{10}$, then at least $T\sim 1 s$. This gives
a lower bound for the length in time of the experiment (in practice it
will be much longer). This leads to a dispersion of the particles. In
order to estimate the dispersion in the length of time considered, we
note that for a Gaussian packet, for a minimum dispersion packet one
has that the dispersion is \cite{cohen},
\begin{equation}
\Delta x(t) =\frac{\delta}{2}\sqrt{1 + \frac{4\hbar^2t^2}{m^2 \delta^4}},\label{dispersion}
\end{equation}
where $\delta$ is a parameter that gives the width of the Gaussian. If we minimize $\delta$ for a given time $T$ one gets $\delta^2 =2\hbar T/m$
and therefore the minimum dispersion is given by,
\begin{equation}
\Delta x \sim \sqrt{\frac{\hbar T}{m}}\sim 10^{-5}m,
\end{equation}
for the example in question. Due to this dispersion we will have very
different velocities for the different values of the impact parameter
$d$ ranging from $v_y\sim 10^6$ m/s to $10^2$ m/s for $d\sim
10^{-13}$m and $10^{-5}$m respectively. These velocity differences give
rise to deviations from the original trajectory one sent the spins
into the cavity without any control of the experimenter. To avoid
these problems and have more or less predictable trajectories for the
spins of the environment one should have to make the interaction
distance much larger than before, let us say $d\sim 10^{-4}m$. This
would lead to too weak an interaction with the spins of the
environment and therefore no decoherence takes place.

One could ask what would happen if one did not use a package which
minimizes uncertainty growth, as we did before.  One can see that in
that case, if one chooses a sufficiently small packet initially to
avoid large deviations in the $y$ direction the size of the packet
grows rapidly with time. That would force to send the spins of the
environment with large separations in time in order to avoid them from
interacting among each other. Allowing out of control interactions
among the environment spins will prevent us from computing the
observable. Using packets that are very spaced in time avoids that
problem initially, but the large size of the packages at the end of
the experiment does not allow us to have particles of the environment
that do not interact with each other, as we discuss in the appendix.

The conclusion is that, at least with nuclear particles, it is
impossible to compute the observable because if one requires enough
interaction among the spins for decoherence to take place, one ends up
losing control of the experiment. That is, one cannot measure the observable.

Could it be that the experiment is still feasible with larger masses and
magnetic momenta? We will analyze this situation in the next section.

\section{An experiment with large masses and magnetic momenta}
\subsection{The experiment and the appearance of decoherence}
\label{Va}
As we mentioned in the previous section, the problems in measuring the
observable could in principle be overcome if one used a larger separation between
the spins of the environment and the spin of the ``needle'', and using larger
magnetic moments. The larger magnetic moment will enable decoherence in
spite of the larger separation. One will still have large deviations from the initial
direction of the trajectory due to the interactions, but since the separation is
large compared to the size of the packet, one will have little dispersion in the
direction of exit. This will require a full quantum treatment of the problem. Let us
go back to the interaction Hamiltonian (\ref{hint}), and rewrite it in terms of the
vectors of the spin basis along the $z$ direction.

Using that
\begin{eqnarray}
\hat{S}_x&=&\vert+\rangle\langle-\vert+\vert-\rangle\langle+\vert\\
\hat{S}_y&=&-i\vert+\rangle\langle-\vert+i\vert-\rangle\langle+\vert\\
\hat{S}_z&=&\vert+\rangle\langle+\vert-\vert-\rangle\langle-\vert
\end{eqnarray}
we can represent the Hamiltonian as,
\begin{equation}\label{haminter}
\hat{H}_k = \left(
\begin{array}{cccc}
f_k+B\Gamma_+&0&0&0\\
0&-f_k+B\Gamma_-&2f_k&0\\
0&2f_k&-f_k-B\Gamma_-&0\\
0&0&0&f_k-B\Gamma_+
\end{array}\right),\label{1.4}
\end{equation}
where the elements of the basis are ordered as $\vert ++\rangle$,
$\vert +-\rangle$, $\vert -+\rangle$, $\vert --\rangle$ respectively
and $\Gamma_\pm\equiv \gamma_1\pm\gamma_2$

With this Hamiltonian we can find the evolution of the system when the
spin $k=1$ is interacting with $S$ (we initially focus in the passage of
the first spin in the stream through the cavity),
\begin{eqnarray}\label{1.5}
i \frac{d\vert \psi\rangle}{dt} &=& \hat{H}_1 \vert\psi\rangle,\\
\vert\psi(t=0)\rangle &=& \left(a\vert+\rangle+b\vert-\rangle\right)\otimes
\label{1.6}
\left(\alpha_1\vert+\rangle_1 +\beta_1 \vert-\rangle_1\right),\\
\vert\psi(t)\rangle&=&R(t)\vert+\rangle\vert+\rangle_1+
T(t)\vert+\rangle\vert-\rangle_1+
U(t)\vert-\rangle\vert+\rangle_1+
V(t)\vert-\rangle\vert-\rangle_1 \label{1.7},
\end{eqnarray}
and using (\ref{1.4}-\ref{1.6}) in (\ref{1.7}) we get a set of differential
equations for the coefficients,
\begin{eqnarray}
i\dot{R}&=& \left(f_1+B\Gamma_+\right) R,\\
i\dot{U}&=& 2f_1 T+ \left(-f_1-B\Gamma_-\right) U,\\
i\dot{T}&=& 2f_1 U+ \left(-f_1+B\Gamma_-\right) T,\\
i\dot{V}&=& \left(f_1-B\Gamma_+\right) V,
\end{eqnarray}
where we have assumed that $f_k$ is time independent. Notice that we
are neglecting a proper treatment of the spatial dependence of the
quantum states. We are carrying out a semiclassical analysis in which
each spin is treated like a classical point particle that flies
through the cavity in a well defined trajectory and such that the
distance between spin and the ``needle'' spin changes little during
its flight in the cavity. The only quantum aspect we are treating is
the spin-spin and spin-field interactions. The treatment is
quantum-statistical mechanics in nature, in the sense that the
couplings $f_k$ must be taken to be a random variable whose
statistical properties will be determined by the spread of the
wavefunction associated with the particle $k$. This is a good
approximation if the distance from spin $k$ to spin $S$ is
approximately constant during the time of flight of $k$ within the
cavity, as would be the case in a thin cavity if $k$ impacts far away
from $S$.  The initial conditions for the differential equations are
$R(t=0)=a\alpha_1$, $T(t=0)=a\beta_1$, $U(t=0)=b\alpha_1$,
$V(t=0)=b\beta_1$. We can solve for $R$ and $V$ and get,
\begin{eqnarray}\label{1.8}
R(t)&=& a \alpha_1 \exp\left(-i\left(f_1+B\Gamma_+\right)t\right),\\
V(t)&=& b \beta_1 \exp\left(-i\left(f_1-B\Gamma_+\right)t\right),
\end{eqnarray}
and for $T$ and $U$,
\begin{eqnarray}
T(t)&=& a \beta_1 e^{if_1 t} \left(\cos\left(\Omega_1 t\right)-i \frac{B\Gamma_-}{\Omega_1}\sin\left(\Omega_1 t\right)\right)-2ib\alpha_1
\frac{f_1}{\Omega_1}e^{if_1t}\sin\left(\Omega_1 t\right),\\
U(t)&=& b \alpha_1 e^{if_1 t} \left(\cos\left(\Omega_1 t\right)+i \frac{B\Gamma_-}{\Omega_1}\sin\left(\Omega_1 t\right)\right)-2ia\beta_1
\frac{f_1}{\Omega_1}e^{if_1t}\sin\left(\Omega_1 t\right),\\\label{1.11}
\end{eqnarray}
where we have introduced the frequency
$\Omega_1=\sqrt{4f_1^2+B^2\Gamma_-^2}$.

Replacing the solutions (\ref{1.8}-\ref{1.11}) in (\ref{1.7}) we have the state
of the system after the passage of spin $k=1$ is given by,
\begin{eqnarray}
\vert\psi(t)\rangle_1 &=&
 a\vert+\rangle
\left[
\alpha_1 e^{-i\left(f_1+B\Gamma_+\right)t} \vert+\rangle_1
+\beta_1e^{if_1 t}
\left(\cos\left(\Omega_1 t\right)-i \frac{B\Gamma_-}{\Omega_1}\sin\left(\Omega_1 t\right)\right)\vert-\rangle_1 \nonumber\right.\\
&&\left.-2 i \frac{b\alpha_1}{a}\frac{f_1}{\Omega_1} e^{if_1 t} \sin\left(\Omega_1 t\right)
\vert-\rangle_1
\right]\\
&&+
b\vert-\rangle
\left[
\beta_1 e^{-i\left(f_1-B\Gamma_+\right)t} \vert-\rangle_1
+\alpha_1e^{if_1 t}
\left(\cos\left(\Omega_1 t\right)+i \frac{B\Gamma_-}{\Omega_1}\sin\left(\Omega_1 t\right)\right)\vert+\rangle_1 \right.\nonumber\\
&&\left.-2 i \frac{a\beta_1}{b}\frac{f_1}{\Omega_1} e^{if_1 t} \sin\left(\Omega_1 t\right)
\vert+\rangle_1
\right],
\end{eqnarray}
and given $\tau$ the time of flight of the spins in the cavity we can
find the state of the system after $N$ spins have passed,
\begin{equation}
\vert\psi\rangle=a\vert+\rangle\vert A\rangle+b \vert-\rangle\vert B\rangle
\end{equation}
with,
\begin{eqnarray}
\vert A\rangle&=&\prod_k^N\left[
\alpha_k e^{-i\left(f_k+B\Gamma_+\right)\tau} \vert+\rangle_k
+\beta_ke^{if_k \tau}
\left(\cos\left(\Omega_1 \tau\right)-i \frac{B\Gamma_-}{\Omega_k}\sin\left(\Omega_k \tau\right)\right)\vert-\rangle_k \nonumber\right.\\
&&\left.-2 i \frac{b\alpha_k}{a}\frac{f_k}{\Omega_k} e^{if_k \tau} \sin\left(\Omega_k \tau\right)
\vert-\rangle_k
\right]\\
\vert B\rangle&=&
\prod_k^N
\left[
\beta_k e^{-i\left(f_k-B\Gamma_-\right)\tau} \vert-\rangle_k
+\alpha_ke^{if_k \tau}
\left(\cos\left(\Omega_k \tau\right)+i \frac{B\Gamma_-}{\Omega_k}\sin\left(\Omega_k \tau\right)\right)\vert+\rangle_k \right.\nonumber\\
&&\left.-2 i \frac{a\beta_k}{b}\frac{f_k}{\Omega_k} e^{if_k \tau} \sin\left(\Omega_k \tau\right)
\vert+\rangle_k
\right].
\end{eqnarray}

The density matrix operator of the system is given by,
\begin{equation}
\hat{\rho} = \vert \psi\rangle \langle\psi \vert=
\vert a\vert^2 \vert+\rangle\langle +\vert \vert A\rangle\langle A\vert +
ab^*\vert+\rangle\langle-\vert\vert A\rangle\langle B\vert+
a^* b\vert-\rangle\langle+\vert\vert B \rangle\langle A\vert +
\vert b\vert^2 \vert-\rangle\langle - \vert \vert B\rangle\langle B\vert,
\end{equation}
and the reduced density matrix of the ``needle gauge'' system $S$ is
\begin{equation}
\hat{\rho}_S =\vert a\vert^2 \vert+\rangle\langle+\vert\langle A\vert A\rangle+
ab^* \vert+\rangle\langle-\vert\langle B\vert A\rangle+
a^*b \vert-\rangle\langle+\vert\langle A\vert B\rangle+
\vert b\vert^2 \vert-\rangle\langle-\vert\langle B\vert B\rangle.
\end{equation}

We now need to show that the crossed terms in the needle basis go to
zero in the large $N$ limit. We have that,
\begin{eqnarray}
\langle A\vert A\rangle&=&\prod_k^N\left[\vert\alpha_k\vert^2+\vert\beta_k\vert^2
\left(\cos^2\left(\Omega_k\tau\right)+\left(\frac{B\Gamma_-}
{\Omega_k}\right)^2\sin^2\left(\Omega_k \tau\right)\right)\right.\nonumber\\
&&
+4\frac{\vert b\vert ^2\vert \alpha_k\vert ^2}{\vert a\vert ^2}\frac{f_k^2}{\Omega_k^2}\sin^2\left(\Omega_k
\tau\right)
+2i\beta_k \left(\frac{b\alpha_k}{a}\right)^*\frac{f_k}{\Omega_k}
\left(\cos\left(\Omega_k\tau\right)-i\frac{B\Gamma_-}{\Omega_k}\sin\left(\Omega_k\tau\right)\right)\sin\left(\Omega_k\tau\right)\nonumber\\
&&\left.-2i\frac{b\alpha_k}{a}\beta_k^*\frac{f_k}{\Omega_k}\left(
\cos\left(\Omega_k\tau\right)+i\frac{B\Gamma_-}{\Omega_k}
\sin\left(\Omega_k\tau\right)\right)\sin\left(\Omega_k\tau\right)\right]
\end{eqnarray}
with a similar expression for $\langle B\vert B\rangle$. For the last term we have,
\begin{eqnarray}
\langle A\vert B\rangle &=&
\prod_k^N \left[
\vert \alpha_k\vert ^2 e^{2if_k\tau+iB\Gamma_+\tau}
\left(\cos\left(\Omega_k\tau\right)+i\frac{B\Gamma_-}{\Omega_k}\sin\left(\Omega_k\tau\right)\right)-
\alpha_k^* e^{i\left(2f_k+B\Gamma_+\right)\tau}
2 i \frac{a\beta_k}{b}\frac{f_k}{\Omega_k}\sin\left(\Omega_k\tau\right)\right.\\
&&\left.+\vert \beta_k\vert ^2 e^{-i\left(2f_k-B\Gamma_+\right)\tau}
\left(\cos\left(\Omega_k\tau\right)+i\frac{B\Gamma_-}{\Omega_k}\sin\left(\Omega_k\tau\right)\right)
+2 i \left(\frac{b\alpha_k}{a}\right)^*\beta_k \frac{f_k}{\Omega_k}
e^{-i\left(2f_k -B\Gamma_+\right)\tau}
\sin\left(\Omega_k \tau\right)\right].\nonumber
\end{eqnarray}

If we now consider the case in which the coupling between the spins is
much weaker than the coupling with the magnetic field, and that $\gamma_1\neq
\gamma_2$ i.e.,$f_k \ll \vert B\Gamma_-\vert $, one has that,
\begin{equation}\label{1.17}
\Omega_k=\sqrt{4f_k^2+B^2\Gamma_-^2}\sim
B\Gamma_-\left(1+\frac{1}{2}\frac{4f_k^2}{B^2\left(\gamma_1-
\gamma_2\right)^2}\right),
\end{equation}
so the leading term is $\Omega_k\sim B\Gamma_-$ and
therefore $f_k/\Omega_k\ll 1$.

Using these approximations in the above expressions for the inner
products,
\begin{equation}
\langle A\vert A\rangle =\prod_k^N \left[\vert \alpha_k\vert ^2+\vert \beta_k\vert ^2
\left(\cos^2\left(\Omega_k \tau\right)+\sin^2\left(\Omega_k \tau\right)\right)\right]
=1,
\end{equation}
with a similar expression for $\langle B\vert B\rangle$.
and
\begin{eqnarray}
\langle A \vert B\rangle &=&\prod_k^N \left[
\vert \alpha_k\vert ^2 e^{2if_k\tau+i\Omega_k\tau}
\left(\cos\left(\Omega_k\tau\right)+i\sin\left(\Omega_k\tau\right)\right)
+\vert \beta_k\vert ^2 e^{-2if_k\tau+i\Omega_k\tau}
\left(\cos\left(\Omega_k\tau\right)+i\sin\left(\Omega_k\tau\right)\right)\right]\\
&=& \prod_k^Ne^{2i\Omega_k\tau}
\left[\cos\left(2f_k\tau\right)+i\left(\vert \alpha_k\vert ^2-\vert \beta_k\vert ^2\right)\sin\left(2f_k\tau\right)\right].
\end{eqnarray}
This last expression goes to zero for large $N$ since it is the product
of complex numbers of unit modulus but with random phases.
We can therefore construct the reduced density matrix,
\begin{eqnarray}
\hat{\rho}_S&=&\vert a\vert ^2 \vert+\rangle\langle+\vert+
ab^*\vert+\rangle\langle-\vert
\left(\prod_k^N e^{2i\Omega_k\tau}
\left(\cos\left(2f_k\tau\right)+
i\left(\vert \alpha_k\vert ^2-\vert \beta_k\vert ^2\right)\sin\left(2f_k\tau\right)\right)\right)^*\nonumber\\
&&+a^*b\vert-\rangle\langle+\vert\prod_k^N e^{2i\Omega_k\tau}
\left(\cos\left(2f_k\tau\right)+
i\left(\vert \alpha_k\vert ^2-\vert \beta_k\vert ^2\right)\sin\left(2f_k\tau\right)\right)+
\vert b\vert ^2\vert-\rangle\langle-\vert,
\end{eqnarray}
so we have decoherence,
\begin{equation}
\hat{\rho}_S=\left(
\begin{array}{cc}
\vert a\vert ^2\langle A\vert A\rangle& ab^*\langle B\vert A\rangle\\
a^*b\langle A\vert B\rangle& \vert b\vert ^2\langle B\vert B\rangle
\end{array}\right)
\begin{array}{c}
\rightarrow\\
{\scriptstyle N\gg 1}
\end{array}
\left(
\begin{array}{cc}
\vert a\vert ^2 & 0\\
0& \vert b\vert^2
\end{array}
\right)
\end{equation}

There is a caveat in this proof of decoherence. For a closed system,
the off diagonal terms are given really by multiperiodic functions,
i.e. they are given by a superposition of a large number of periodic
functions with different frequencies. Therefore this function will
retake values arbitrarily close to the initial value for sufficiently
large times. This implies that the apparent loss of information about
the non-diagonal terms reappears if one waits a long enough time. This
problem is usually called ``recurrence of coherence'' (``revivals'').
The characteristic time for these phenomena is proportional to the
factorial of the number of involved frequencies. 

The above derivation was done using ordinary quantum mechanics in
which one assumes an ideal clock is used to measure time. As we have
argued in our previous paper, If one redoes the derivation using the
effective equation we derived for quantum mechanics with real clocks
one gets the same expression for the off-diagonal terms except that it
is multiplied by $\prod_k\exp\left(-(2 g_k)^2 T^{4/3}_{\rm Planck}
t^{2/3}\right)$.  That means that asymptotically the off diagonal
terms indeed vanish, the off-diagonal terms are not periodic
anymore. Although the exponential term decreases slowly with time, the
fact that there is a product of them makes the effect quite relevant,
especially for the long periods of time involved in
``revivals''. Therefore we see that including real clocks in quantum
mechanics offers a mechanism to turn pure states into mixed states in
a way that is desirable to explain the problem of measurement in
quantum mechanics.

How many particles does one need to consider for the exponential
decrease to kill the possibility of revivals? A criterion would be that
the magnitude of the off diagonal term in the revivals be smaller than
the magnitude of the off diagonal terms in the intermediate region between
revivals. If that were the case the revival would be less than the
``background noise'' in regions where there is no revival. The magnitude
of the interference terms in the density matrix were studied by
Zurek  \cite{zurek} and go as $\rho_{+-} \sim 1/2^{N/2}$. The time
for revivals to occur goes as $T\sim N!/\Omega$ where $\Omega$ is
the mean value of the $\Omega_k$'s. This implies that if one
has more than hundreds of particles the loss of coherence will make
impossible the observation of revivals.

At this point it is worthwhile emphasizing the robustness of this result
in practical terms. One could, for instance, question how reliable the
fundamental limit for the inaccuracy of clocks of \cite{ng} is. Some
authors have characterized the fundamental limit as too optimistically
large, arguing that the real fundamental limit is of the order of Planck
time itself. In view of this it is interesting to notice that if one posits a
much more conservative estimate of the error of a clock, for instance
$\delta T \sim T^\epsilon T_{\rm Planck}^{1-\epsilon}$, for any small
value of $\epsilon$ the only modification would be to change the
number of particles $N_0\sim 100$ to at least $N\sim N_0/(3\epsilon)$.

\subsection{Non-observability of d'Espagnat's global observable}

\label{Vb}
As we mentioned above, d'Espagnat proposed a global observable for
Zurek's model that could be used to test the presence of decoherence
versus state reduction in the model we are considering. Let us analyze
how the observable behaves in the latter. It is defined as,
\begin{equation}
\hat{M} \equiv \hat{S}_x \otimes\prod_k^N \hat{S}^k_x.
\end{equation}

Let us assume that at some point in the experiment the wavefunctions
collapses. When the collapse occurs we would have to project the
system onto the final state. For instance, if we assume the measured
spin is up we would have $a=1,b=0$. From there on the system would
continue its evolution, as discussed above. The important point is that
the evolution considered will not move the system away from its
collapsed state, that is $a$ would still be unity after evolution. That
is, from the moment the collapse happens we would have $\langle\hat{M}\rangle=0$.

Let us go back and consider that no collapse occurs and compute
the evolution of the expectation value of the observable.

{}From the approximations (\ref{1.17}) one has that the state is,
\begin{equation}
\vert \psi\rangle =
a \vert +\rangle
\prod_k^N \left[
\alpha_k e^{-i\left(f_k+\Omega_k\right)\tau} \vert +\rangle_k
+\beta_k e^{i\left(f_k-\Omega_k\right)\tau} \vert -\rangle_k\right]
+
b \vert -\rangle
\prod_k^N \left[
\alpha_k e^{-i\left(f_k+\Omega_k\right)\tau} \vert +\rangle_k
+\beta_k e^{-i\left(f_k-\Omega_k\right)\tau} \vert -\rangle_k\right]
\end{equation}
so the expectation value of the observable\footnote{The reader
may ponder if this observable is measurable even in principle, since
it is generically very small since it involves large products of small
factors. However, one could in principle prepare the initial state
in such a way that the products are not small, for instance choosing
$\alpha_k=\beta_k=1/\sqrt{2}$.}
 is,
\begin{equation}
\langle \psi\vert M\vert\psi\rangle =
ab^*\prod_k^N\left[\alpha_k\beta_k^* +\alpha^*_k\beta_k\right]
e^{-2i\Omega_k\tau} + a^*b
\prod_k^N\left[\alpha_k\beta_k^* +\alpha^*_k\beta_k\right]
e^{2i\Omega_k\tau},
\end{equation}
and the density matrix for the pair $S,S_k$, is given by,
\begin{equation}
\rho=\left(
\begin{array}{cccc}
\vert a \vert^2 \vert \alpha_k \vert^2 &
\vert a \vert^2 \alpha_k \beta_k^*&
ab^* \vert\alpha_k\vert^2 &
a b^* \alpha_k \beta_k^*\cr
\vert a \vert^2 \alpha_k^* \beta_k&
\vert a \vert^2 \vert \beta_k \vert^2 &
a b^* \alpha^*_k \beta_k &
a b^* \vert\beta_k\vert^2 \cr
a^*b \vert\alpha_k\vert^2 &
a^* b \alpha_k \beta_k^*&
\vert b\vert^2 \vert\alpha_k\vert^2&
\vert b\vert^2 \alpha_k \beta_k^*\cr
a^* b \alpha_k^* \beta_k &
a^* b \vert \beta_k\vert^2&
\vert b\vert^2 \alpha_k^* \beta_k &
\vert b\vert^2 \vert \beta_k \vert^2
\end{array}\right).
\end{equation}

The expression for the state clearly differs from the one obtained in the case with
collapse. For instance, taking into account that we are working in the
limit $\Omega_k\sim B\Gamma_-$ and therefore $\Omega$ is
$k$-independent, and choosing the time of flight $\tau$ such that
$e^{\pm 2i\Omega N \tau}=1$ one has that $\langle M\rangle$ is
obviously different from zero.

As we discussed above, an experiment measuring this type of global
observables is not possible using nuclear particles. However, for
quantum systems with lager masses and magnetic momenta it is clearly
possible to measure them. One could therefore distinguish collapse
from reduction in the wavefunction. We will see that the fundamental
loss of coherence due to the use of realistic clocks and rods eliminates
the possibility of measuring the global observable.

Let us consider the evolution of the observable when we consider
the modified evolution due to real rods and clocks. In that case we have
shown \cite{obregon} that the off-diagonal terms of the density matrix
in the energy pointer basis decrease exponentially as,
\begin{equation}
\rho_{mn}(T) = \rho_{mn}(0) \exp\left(i\omega_{mn} T\right)
\exp\left(-\omega_{mn} T_{\rm Planck}^{4/3} T^{1/3}\right)
\end{equation}
with $a>0$ and $\omega_{mn}$ are the Bohr frequencies between the
levels $n$ and $m$. The relevant Bohr frequencies for this case can
be obtained by considering
consider the interaction Hamiltonian  (\ref{haminter}) in the approximation
where $f_k <\vert B \gamma_{1,2}\vert$,
\begin{equation}
\hat{H}_k = {\rm diag} \left(
B\left(\gamma_1 +\gamma_2\right),
B\left(\gamma_1 -\gamma_2\right),
-B\left(\gamma_1 -\gamma_2\right),
-B\left(\gamma_1 +\gamma_2\right)\right).
\end{equation}
The end result is that the evolution of the density matrix is given by,
\begin{equation}
\rho=\left(
\begin{array}{cccc}
\vert a \vert^2 \vert \alpha_k \vert^2 &
\vert a \vert^2 \alpha_k \beta_k^* e^{-(2 B \gamma_2)^2\theta}&
ab^* \vert\alpha_k\vert^2 e^{-(2 B \gamma_1)^2\theta}&
a b^* \alpha_k \beta_k^* e^{-(2 B (\gamma_1+\gamma_2))^2\theta}\cr
\vert a \vert^2 \alpha_k^* \beta_k e^{-(2 B \gamma_2)^2\theta}&
\vert a \vert^2 \vert \beta_k \vert^2 &
a b^* \alpha^*_k \beta_k e^{-(2 B (\gamma_1-\gamma_2))^2\theta}&
a b^* \vert\beta_k\vert^2 e^{-(2 B (\gamma_1))^2\theta}\cr
a^*b \vert\alpha_k\vert^2 e^{-(2 B \gamma_1)^2\theta}&
a^* b \alpha_k \beta_k^*e^{-(2 B (\gamma_1-\gamma_2))^2\theta}&
\vert b\vert^2 \vert\alpha_k\vert^2&
\vert b\vert^2 \alpha_k \beta_k^*e^{-(2 B (\gamma_2))^2\theta}\cr
a^* b \alpha_k^* \beta_k e^{-(2 B (\gamma_1+\gamma_2))^2\theta}&
a^* b \vert \beta_k\vert^2e^{-(2 B (\gamma_1))^2\theta}&
\vert b\vert^2 \alpha_k^* \beta_k e^{-(2 B (\gamma_2))^2\theta}&
\vert b\vert^2 \vert \beta_k \vert^2
\end{array}\right)
\end{equation}
where $\theta=3 T_{\rm Planck}^{4/3} \tau^{2/3}/2$.

A calculation very
similar to the one we did before but with the modified evolution yields,
\begin{eqnarray}
\langle \hat{M}\rangle &=&
a b^* e^{-i2 N \Omega T} e^{-4 N B^2 (\gamma_1-\gamma_2)^2 \theta}
\prod_{k}^N \left[ \alpha_k \beta_k^* e^{-16 B^2\gamma_1\gamma_2\theta}
+\alpha_k^* \beta_k\right]\\
&&+b a^*
e^{i2 N \Omega T} e^{-4 N B^2 (\gamma_1-\gamma_2)^2 \theta}
\prod_{k}^N \left[ \alpha_k \beta_k^*
+\alpha_k^* \beta_ke^{-16 B^2\gamma_1\gamma_2\theta}\right]
\end{eqnarray}
where $\Omega= B(\gamma_1 -\gamma_2)$. We therefore see that
the expectation value of the observable decreases exponentially with
time with an exponent proportional to $B (\gamma_1 -\gamma_2)$.
This is  a large number since as we noted $B (\gamma_1 -\gamma_2)>f_k$
and the $f_k$'s are large as we argued above.

Let us study if there is a range of parameters where at least in
principle, one could measure the observable and therefore conclude
that collapse has occurred or not. Let us summarize the set of conditions we
have encountered in this work: a) that the couplings be large enough
so decoherence can occur, b) that there is no significant dispersion
in the particles, c) that the interaction with the magnetic field be
larger than the inter-spin interaction so one has decoherence in a
predetermined basis, d) the effects implied by the use of real rods and
clocks in measurement. These correspond to:
\begin{eqnarray}
&&a) \qquad 1< f\tau =\frac{\mu\gamma_1 \gamma_2}{\hbar}\frac{\tau}{d^3},
\label{a}\\
&&b) \qquad \Delta x \sim \sqrt{\frac{\hbar T}{m}}, \label{b}\\
&&c)\qquad f\ll \vert B(\gamma_1 -\gamma_2)\vert, \label{c}\\
&&d)\qquad <\hat{M}> \sim \exp\left(-6 N B^2(\gamma_1-\gamma_2)^2 T_{\rm Planck}^{4/3} \tau^{2/3}\right),\label{d}
\end{eqnarray}
where $T$ is the total length of the experiment and $\tau$ the time of
flight within the cavity. Let us consider (\ref{b}) with the condition
$T> N\tau$ since the particles are assumed to enter the cavity one at
a time, then,
\begin{equation}
\Delta x > \sqrt{\frac{\hbar N \tau}{m}}. \label{5.5}
\end{equation}
Now, from (\ref{a}) one has that,
\begin{equation}
d^3 < \frac{\mu\gamma_1\gamma_2}{\hbar}\tau.\label{5.6}
\end{equation}
In addition to that we need that $\Delta x <d$ otherwise, i) if the
dispersion were larger than the distance between the environment
particles and the needle, we could not avoid collisions among them,
ii) the deflection due to the magnetic interaction would be very large
if the particles could impact arbitrarily close (as would happen if
the dispersion were larger than the impact parameter), iii) condition
(\ref{c}) would be violated if the particles flew by closely, since
the coupling would become large. Combining (\ref{5.5}) and (\ref{5.6})
we have that,
\begin{equation}\label{5.7}
\tau^{1/3} < \frac{m(\gamma_1\gamma_2)^{2/3}}{h^{5/3} N} \mu^{2/3}.
\end{equation}
Considering now the exponent in (\ref{d}) $K = N B^2 (\gamma_1-\gamma_2)^2 T_{\rm Planck}^{4/3} \tau^{2/3}$ we have that, using (\ref{a}) and (\ref{b}),
\begin{equation}
B^2(\gamma_1-\gamma_2)^2\tau^2 \gg f^2 \tau^2>1
\end{equation}
and therefore
\begin{equation}
K = \frac{N B^2(\gamma_1-\gamma_2)^2 \tau^2 T_{\rm Planck}^{4/3} \tau^{2/3}}
{\tau^2} \gg \frac{N T_{\rm Planck}^{4/3}}{\tau^{4/3}},
\end{equation}
and replacing in this expression (\ref{5.7}) we have that,
\begin{equation}
K \gg \frac{N T_{\rm Planck}^{4/3}}{\tau^{4/3}}>
\frac{NT_{\rm Planck}^{4/3} \hbar^{20/3} N^4}{m^4 (\gamma_1 \gamma_2)^{8/3} \mu^{8/3}}.
\end{equation}
Let us recall that if $K$ is large we will not be able to decide if
the system underwent collapse or not. We therefore have to impose that
$K<1$ in order to distinguish if there was collapse or not. From the
previous expression this implies that,
\begin{equation}
m (\gamma_1 \gamma_2)^{2/3} \gg \frac{T_{\rm Planck}^{1/3} \hbar^{5/3} N^{5/4}}{\mu^{2/3}}.
\end{equation}
Let us see if we can meet that condition. Taking the ``needle'' spin
as a proton and environment spins to be neutrons (the particles have to be
different, see \footnote{The
particles need to have different magnetic moments in order for an
external magnetic field to determine the pointer basis. This is
reflected in the $\gamma_1-\gamma_2$ factors that appear in various
expressions. Decoherence still takes place due to the interaction of
the needle and the environment, but there is not a preferred basis in
which the events will occur.}) one gets that $10^6\gg N^{5/4}$, which
is violated already for an ``environment'' with $10^5$ spins. We know
this case is already not feasible due to the interactions and the
dispersions, but it is good to verify that even if one did not have
those difficulties, the fundamental loss of coherence would prevent us
from deciding that collapse has occurred.

It is worthwhile asking what would happen if one considered particles
with significantly more mass and magnetic moment, to avoid the issues
of the close interaction and associated dispersion. Let us assume that we consider an environment of $10^{23}$ spins. one has that,
\begin{equation}
m (\gamma_1 \gamma_2)^{2/3} > 10^{-38}
\end{equation}
which requires masses and magnetic moments much larger than those of
elementary particles. For instance if we consider particles with
$m=M_{\rm Planck}$ the gyromagnetic ratios needed are of the order of
$10^{20}$ and that have to live at least $10^{-4}s$. Such objects are
highly unlikely to exist due to quantum field theory effects.

\section{Discussion}

We have argued that fundamental limitations in the process of
measurement, both due to gravitational and quantum mechanical effects
lead to a loss of coherence in quantum evolution. Coupled to
environmentally-induced decoherence, we conjecture that this effect
provides a solution to the problem of measurement in quantum
mechanics. Basically, we propose that the collapse or not of a state
becomes undecidable and when that happens an event has taken place.
In this paper we have examined these ideas in the context of models
that are usually considered when discussing decoherence. We have
shown that indeed the collapse or not of the states becomes undecidable,
and that using undecidability to characterize when an event has taken
place solves the ``problem of outcomes''. We have also studied the
possibility of using global observables, as proposed by d'Espagnat,
to characterize if collapse has taken place and we have shown that
the limitations in measurement we point out make impossible the
measurement of the relevant global observables.

The reader may question how distinguishable are the fundamental
limitations we are discussing in this paper from the practical
limitations that have often been invoked when discussing the solution
of the problem of measurement through environmentally-induced
decoherence.  From our point of view the key difference is the
exponential decay of observables due to fundamental limitations in
measurement. For instance, in the example we consider, to distinguish
an expectation value that is exponentially small and one that is
identically zero would require a measurement of an ensemble with a
number of elements that quickly becomes prohibitive, say, comparing to
the total degrees of freedom of the universe inside our horizon. This
limitation bears some resemblance to the non-polynomial computations
in quantum computing. In that case, an NP problem cannot be worked out
no matter what the details of the computer in question. In our case,
the effect we discuss will occur irrespective of the details of the
experiment in question and the particulars of the mechanism for
fundamental loss of coherence, and is essentially limited by the size
of the universe.

To conclude we point out that our results have been derived in a
particular model, and further work in better models is needed to build
a stronger case that the resolution of the problem of measurement
presented is a robust one.

\section{Acknowledgments}

We wish to thank Mario Castagnino and Olimpia Lombardi for
discussions.  LPGP would like to dedicate this work to Dr Alberto
Barcia, for being always an inspiration. This work was supported in
part by grant NSF-PHY-0650715, funds of the Hearne Institute for
Theoretical Physics, FQXi, CCT-LSU, Pedeciba and ANII PDT63/076.

\section*{Appendix 1}

Here we would like to discuss what happens if one attempts to use
packets that do not minimize the uncertainty growth. These would be
packets that start small and grow over time. We shall see that the growth
is too large in the end, leading to packets that cannot be considered
non-interactive and therefore for which one cannot compute the global
observables.

The speed that the particles acquire due to the magnetic interaction
is given by the condition (\ref{vy}) $v_y \ge \hbar/(m d)$ and the
dispersion as a function of time is (\ref{dispersion}),
\begin{equation}
\Delta x(t) = \frac{\delta}{2}\sqrt{1+\frac{4\hbar^2 t^2}{m^2 \delta^4}}.
\end{equation}
We will choose the initial width of the packet in such a way that they
do not suffer very different deviations in their trajectories due to
the magnetic interaction. This way, even though they deviate, they
will all do it in approximately the same way and we know where to find
them at the end of the experiment. If we chose the dispersion of the order
of the impact parameter some particles would come very close to the
needle and would deviate a lot. We will take $\delta=d/10$, that is one order
of magnitude smaller than the impact parameter and we will study if the
experiment is feasible. The dispersion at the end of the experiment
therefore is,
\begin{equation}
\Delta x(T)= \frac{d}{20} \sqrt{1 + \frac{4 \hbar^2 T^2}{m^2 (0.1 d)^4}}
\ge \frac{\hbar T}{0.1 m d}.\label{69}
\end{equation}
The condition to have loss of coherence (sufficiently strong couplings)
was given by $1< \mu\gamma_1 \gamma_2/(\hbar d^2 v)$, so we can conclude for the speed that $v<\mu\gamma_1 \gamma_2/(\hbar d^2)$. The total length to be traversed in the experiment can be estimated as
\begin{equation}
l=vT <\mu\gamma_1 \gamma_2 T/(\hbar d^2). \label{70}
\end{equation}

Let us now recall that the final dispersion, caused by the force
(\ref{force}) and the spread of the wave packet in all directions, will
be present not only in the transverse direction to the motion but in
the longitudinal one as well. We therefore have to ensure that the
particles of the environment do not end up interacting among
themselves. If the distance that the particles traverse would be
smaller than the final dispersion the particles could interact. It
would even lead to a non-vanishing probability of finding particles of
the environment in the cavity after the experiment is over. We will
therefore impose $\Delta x(T)<l$.  Using (\ref{69}) and (\ref{70}) we have,
\begin{equation}
\frac{\hbar T}{0.1 m d} \le \Delta X(T) \le l \le \frac{\mu \gamma_1 \gamma2}{\hbar} \frac{T}{d^2},
\end{equation}
which leads to $d\le 0.1 \mu \gamma_1 \gamma_2 m/\hbar^2$. This, for
neutrons, implies $d< 10^{-19}$m, which cannot be satisfied for known
elementary particles. What is going on is that the particles disperse
faster than the distance they are traveling if we impose that they
travel slow enough in order for decoherence with the environment to
take place. Clearly for more massive particles or with larger magnetic
moments one could avoid this problem, but as we discussed in section
\ref{Vb}, the effect of the fundamental decoherence renders the
observable unmeasurable in that case.

\section*{Appendix 2}
\subsection{Events}

Let us briefly outline the formal structure that underlies the notion of event
in our interpretation of quantum mechanics.

Let $\vert \Psi\rangle$ be the resulting state of the evolution of a system
$S$ that has interacted with a measuring device $A$ and an environment
$E$. It is therefore the state of the system ${\bf S}=S\otimes A\otimes E$.
Events generically occur in very complex systems that include a lot of 
elementary subsystems. To simplify the presentation, we will assume that
the system ${\bf S}$ is composed of three spins. A possible state could be,
for instance,
\begin{equation}
\vert \Psi \rangle = 
\frac{c_1}{\sqrt{2}} \left(\vert +,+,-\rangle +\vert +,-,+\rangle\right)
+
{c_2} \vert -,+,+\rangle
\end{equation}
with $\vert c_1\vert^2+\vert c_2\vert^2=1$. In this greatly simplified
picture we will assume $S\otimes A$ is the first spin and $E$ is 
the other two.

The reduced density matrix $\rho$ is the state of the subsystem
$S\otimes A$. It is obtained taking the partial trace over the states
of the unobserved system $E$,
\begin{equation}
\rho ={\rm Tr}_{23}\left(\vert \Psi \rangle \langle \Psi \vert\right) = 
\vert c_1\vert^2 \vert +\rangle\langle +\vert 
+ \vert c_2\vert^2 \vert -\rangle\langle - \vert.
\end{equation}
This density matrix, as was extensively discussed by d'Espagnat,
admits two interpretations. One of them is that $\rho$ is an improper
mixture, since it was obtained as a partial trace of a pure state.  It
represents partial information about a system which we are ignoring a
portion of. $\rho$ represents the options ``spin up'' and ``spin
down'' in coexistence as potentialities with probabilities $\vert
c_1\vert^2$ and $\vert c_2\vert^2$. The same mathematical expression
can be used to describe a situation where the particle was left in the
state ``up'' with probability $\vert c_1\vert^2$ {\em or} with state
``down'' with probability $\vert c_2\vert^2$. Equivalently, it can be
taken to describe a system where a number $\vert c_1\vert^2$ of
particles were introduced in ``up'' state and $\vert c_2\vert^2$ were
introduced in ``down'' state. The ``problem of outcomes'' in quantum
mechanics is how to pass from the first interpretation to the
second. Our proposal is that one should shift the interpretation when
one cannot distinguish the state $\vert \Psi\rangle$ of the whole
system ${\bf S}$ and a statistical superposition of its projections,
\begin{equation}
\rho = \vert c_1\vert^2 \vert \Psi_+\rangle\langle \Psi_+\vert +
\vert c_2\vert^2 \vert \Psi_-\rangle\langle \Psi_-\vert, 
\end{equation}
with,
\begin{eqnarray}
\vert \Psi_+\rangle &=& \vert +\rangle\langle +\vert \otimes I_2\otimes I_3\vert \Psi\rangle\\
\vert \Psi_-\rangle &=&
\vert -\rangle\langle -\vert \otimes I_2\otimes I_3\vert \Psi\rangle,
\end{eqnarray}
since in such a situation any empirical distinction between both
situations is impossible. We say that {\em undecidability} has taken
place. In simple models like the above this does not happen, 
those states are always distinguishable. In systems with many more
degrees of freedom, the fundamental loss of coherence due to gravity
we introduce turns the state indistinguishable from its projections.
That is how undecidability develops, in a universal fashion.

The event implies that the system is now represented by exclusive
alternatives and one of them is adopted by the system. Which
alternative is realized is not predicted by physics, only the
probabilities with which they occur. The chosen alternative is
represented by a the projector $P_{123}$ with eigenvectors $\vert
\Psi_+\rangle$ (or $\vert \Psi_-\rangle$)
\begin{equation}
P_{123} = \frac{1}{2} \left(
\vert+,+,-\rangle 
+\vert+,-,+\rangle \right)
\left(
\langle+,+,-\vert 
+\langle+,-,+\vert \right)
\end{equation}
and all the properties  compatible with it. Those are the properties
associated with subsystems with projectors $P_i$ such that 
$P_i P_{123} = P_{123}$ (they also commute with $P_{123}$).

For example if the property associated to the $\vert \Psi_+\rangle$
state is updated, a compatible property is given by the projector,
\begin{equation}
P_1 = \vert +\rangle\langle +\vert \otimes I_2\otimes I_3,
\end{equation}
which satisfies $P_1 P_{123}=P_{123}$ and characterizes the property
``spin 1 is up''. Another compatible property has projector,
\begin{equation}
P_{23} = I_1\otimes \frac{1}{2}
\left(\vert +,-\rangle+\vert -,+\rangle\right)
\left(\langle +,-\vert+\langle -,+\vert\right)
\end{equation}
which satisfies $P_{23} P_{123}=P_{123}$ and
represents ``spins 2 and 3 are opposite''. 

A quantum event is, in this interpretation, a bundle of properties. 
There is one of them that covers the whole content of the event, in this
case $P_{123}$. This property is in general not accessible 
experimentally given its complexity
We call this property the {\em essential property} of the event since 
it characterizes it completely. All other properties are defined by 
projectors $P_i$ such that $P_i P_{\rm essential} = P_{\rm essential}$.

One last point is that the above discussion could be carried out without 
a measuring apparatus, just with the quantum system $S$ and the 
environment $E$, as we emphasized earlier in the paper.

\end{document}